\documentclass[aps,prd,reprint,amsmath,amssymb,showpacs,superscriptaddress]{revtex4-1}
\usepackage[colorlinks,linkcolor=blue,urlcolor=blue,anchorcolor=blue,citecolor=blue]{hyperref}%
\usepackage{graphicx}
\usepackage{slashed}
\usepackage{subfigure}
\usepackage{bm}
\usepackage{color,xcolor}
\usepackage{dcolumn}
\usepackage{epstopdf}
\usepackage{mathrsfs}
\usepackage{changes}

\begin{document}
	\title{Chiral crossover transition from the Dyson-Schwinger equations in a sphere }
	
	\author{Yin-Zhen Xu}
	\email{xuyz@smail.nju.edu.cn}
	\affiliation{Department of physics, Nanjing University, Nanjing 210093, China}
	
	\author{Chao Shi}
	\email{shichao0820@gmail.com}
	\affiliation{Department of Nuclear Science and Technology, Nanjing University of Aeronautics and Astronautics, Nanjing 210016, China}
	
	\author{Xiao-Tao He}
	\email{hext@nuaa.edu.cn}
	\affiliation{Department of Nuclear Science and Technology, Nanjing University of Aeronautics and Astronautics, Nanjing 210016, China}
	
	\author{Hong-Shi Zong}
	\email{zonghs@nju.edu.cn}
	\affiliation{Department of physics, Nanjing University, Nanjing 210093, China}
	\affiliation{Department of physics, Anhui Normal University, Wuhu 241000, China}
	\affiliation{Nanjing Proton Source Research and Design Center, Nanjing 210093, China}
	\affiliation{Joint Center for Particle, Nuclear Physics and Cosmology, Nanjing 210093, China}
	
	
	\date{\today}
	
	\begin{abstract}
		Within the framework of Dyson–Schwinger equations of QCD, we study the effect of finite volume on the chiral phase transition in a sphere with the MIT boundary condition.  We find that the chiral quark condensate $\langle\bar{\psi} \psi\rangle$ and  pseudotransition temperature $T_{pc}$ of the crossover  decreases as the volume decreases, until there is no chiral crossover transition at last. We find that the system for $R = \infty $\ fm is indistinguishable from $R=10$ fm and there is a significant decrease in $T_{pc}$ with $R$ as $R<4$ fm. When $R<1.5$ fm, there is no chiral transition in the system.
	\end{abstract}
	
	
	\maketitle
	
	\section{Introduction}
	The quantum chromodynamics (QCD) as a underlying theory describing strong interactions exhibits two fascinating aspects: confinement and dynamical chiral symmetry breaking (DCSB). As the temperature increases, the strongly-interacting matter will undergo a phase transition from hadronic matter to quark-gluon plasma (QGP) with deconfinement and chiral restoration. QCD phase transitions are experimental and theoretical frontiers, which have been studied in relativistic heavy-ion collisions (HICs) at CERN (France/Switzerland), BNL (USA), and GSI (Germany) \cite{Adams2005,Shuryak2009}.  However, most of the theoretical calculations are based on the thermodynamic limit (namely, the volume of the system $V \rightarrow \infty$). It is worth bearing in mind that the QGP system produced in HICs always has a finite volume, depending on the collision nuclei, the center of mass energy and the centrality. According to the the UrQMD transport approach \cite{Bass1998}, the volume of Au-Au and Pb-Pb collisions before freeze-out is about $50 \sim 250$ fm$^3$ \cite{Graef2012}. 
	It is believed that  the radii of possible quark gluon plasma are estimated to be 2 to 10 fm. Therefore, there's a problem we need to consider: does the size and shape of QGP system produced in HICs affect the phase transition?\par
	There has been a lot of theoretical studies for the effect of finite volume on QGP phase transition, within the Nambu-Jona-Lasinio (NJL) model \cite{Wang2018,Abreu2019,xia2019}, Polyakov- Nambu-Jona-Lasinio (PNJL) model \cite{Bhattacharyya2013,Pan2017,Liu2020}, quark-meson model \cite{Tripolt2014, Klein2017,Magdy2019}, and Dyson-Schwinger equations (DSEs) \cite{Luecker2010,Li2019,Shi2018,Shi2018a,Shi2020,Zhao2019}. In most existing theoretical studies, for the sake of convenience, the systems are usually treated as a cube, and anti-periodic boundary condition (APBC) is used. However, when the volume of the fireball produced in collision is small enough, not only its size but also its shape have an non-negligible effect on the QCD phase transition. 
	In order to simulate more realistic condition such as the fireball expected to arise in HICs, the authors of Ref. \cite{Zhang2020} consider a sphere with the MIT boundary condition under the framework of NJL model for the first time. However, it should be pointed out that the NJL model is a non-renormalizable theory, in which the confinement property is not preserved. Meanwhile, the gauge sector of QCD, i.e., the gluon degrees of freedom, is lost.
 This led us to consider a more realistic approach to study QGP phase transitions in a sphere with the MIT boundary condition. \par
	In this work, we employ the framework of Dyson-Schwinger equations to deal with the finite size effects in a sphere. DSEs has been widely used in studying strongly interacting phenomena in vacuum and in heat bath \cite{ROBERTS2000S1,RevModPhys.82.2949,Maris:1997tm,Maris:1999nt,Fischer:2014cfa,Hilger:2015ora,PhysRevC.71.065204,PhysRevC.77.025203,PhysRevC.61.045202,Roberts2008,Xu2019,Qin:2010nq}. It is capable of simultaneously implementing color confinement and expressing DCSB \cite{Roberts2008,Roberts1994,Roberts2007,McLerran2007,Bashir:2012fs,Chang:2011vu,Maris:2003vk}. Recently, as mentioned before, it is used to study the effects of finite volume on QGP crossover transition. However, those studies were all for cubic systems i.e. APBC. In this work, we study finite size effects with MIT boundary condition for the first time.  \par
	This paper is organized as follows: In Section \ref{DS}, we introduce the DSEs at finite temperature within the MIT boundary condition. In Section \ref{QC}, chiral quark condensation and chiral susceptibility of a spherical system at different radii are defined and calculated. On this basis, the influence of system volume and shape on chiral crossover transition temperature is discussed, and the results are compared with those obtained by traditional APBC based on cubic systems. In Section \ref{end}, we present a brief summary.
\newpage
\section{Dyson–Schwinger equations  in a finite volume }
\label{DS}
The formulation of DSEs at nonzero temperature are
described in Refs. \cite{ROBERTS2000S1,Bashir:2012fs}. The $T\neq0$ dressed-quark propagator is obtained from
the following gap equation:\\
\begin{align}
{S\left(\tilde{\omega}_{n}, \vec{p}\right)^{-1}}  &{=Z_{2}\left(i \vec{\gamma} \cdot \vec{p}+i \gamma_{4} \tilde{\omega}_{n}+Z_{m} m\right)}\nonumber\\  &{+Z_{1} T \sum_{l} \int \frac{d^{3} p}{(2 \pi)^{3}} g^{2} D_{\mu \nu}\left(k_{\Omega}\right) \frac{\lambda_{a}}{2} \gamma_{\mu} S\left(\tilde{\omega}_{l}, \vec{q}\right) \frac{\lambda_{a}}{2} \gamma_{\nu}},
\label{DSE}　　
\end{align}
where $ \tilde{\omega}_{n}=(2 n+1) \pi T$ are the fermionic Matsubara frequencies; $m$ is the
current-quark mass and $m = 0$ defines the chiral limit. $Z_{1,2,m}$ are the vertex, quark field, and mass renormalization constants, respectively. As we
employ an ultraviolet-finite model, renormalization is unnecessary, i.e. $Z_{1,2,m}=1$. $ D_{\mu \nu}\left(k_{\Omega}\right)$ is the dressed-gluon propagator which has the form:
\begin{align}
g^{2} D_{\mu \nu}\left(k_{\Omega}\right)=P_{\mu \nu}^{T}\left(k_{\Omega}\right) \mathcal{D}\left(k_{\Omega}^{2}\right)+P_{\mu \nu}^{L}\left(k_{\Omega}\right) \mathcal{D}\left(k_{\Omega}^{2}+m_{g}^{2}\right),
\end{align}
where $k_{\Omega}=\left(\tilde{\omega}_{n}-\tilde{\omega}_{l}, \vec{p}-\vec{q}\right)$, and $m_g=(16 / 5) \pi^{2} T^{2}$ is the gluon Debye mass. Since the temperature breaks the Lorentz symmetry, the tensor structure of gluon has both transverse and longitudinal parts, where $ P_{\mu \nu}^{T, L}$ are  transverse and longitudinal projection operators respectively,
\begin{align}P_{\mu \nu}^{T}\left(k_{\Omega}\right) :=\left\{\begin{array}{ll}{0,} & {\mu \text { and/or } \nu=4} \\ {\delta_{i j}-\frac{k_{i} k_{j}}{k^{2}},} & {\mu, \nu=i, j=1,2,3}\end{array}\right.,\end{align}
\begin{align}
P_{\mu \nu}^{L}\left(k_{\Omega}\right)+P_{\mu \nu}^{T}\left(k_{\Omega_{k}}\right)=\delta_{\mu \nu}-\frac{k_{\mu} k_{\nu}}{k^{2}}.
\end{align}
The choice of interaction kernel is not unique. In this work, we use a simplified form of Maris-Tandy model \cite{Qin:2010nq}:
\begin{align}
\mathcal{D}\left(k_{\Omega}^{2}\right)=D_{0} \frac{4 \pi^{2}}{\omega^{6}} k_{\Omega}^{2} e^{-k_{\Omega}^{2} / \omega^{2}}.
\end{align}
The parameters $D_0$ and $\omega$ are not
independent: A change in $D_0$ can be compensated by an
alteration of $\omega$ \cite{Maris:2003vk}. In this paper we choose a typical value $\omega=0.5$ GeV with $D_0\omega=(0.8\ \mathrm{GeV})^{3}$ \cite{Qin:2010nq}.
\par
The gap equation's solution can be generally expressed as
\begin{align}
S\left(\tilde{\omega}_{n}, \vec{p}\right)^{-1}&= i\vec{\gamma} \cdot \vec{p} A\left(\vec{p}^{2}, \tilde{\omega}_{n}^{2}\right)+ B\left(\vec{p}^{2}, \tilde{\omega}_{n}^{2}\right)\nonumber \\ &+i \gamma_{4} \tilde{\omega}_{n} C\left(\vec{p}^{2}, \tilde{\omega}_{n}^{2}\right)+\vec{\gamma} \cdot \vec{p} \gamma_{4} \tilde{\omega}_{n} D\left(\vec{p}^{2}, \tilde{\omega}_{n}^{2}\right),
\end{align}
with the four scalar dressing functions A, B, C, D. The dressing function D, however, is power-law suppressed in the UV \cite{ROBERTS2000S1}
and does not contribute in all cases investigated here. \par
The mass function of quarks can be defined as \cite{Muller:2011awa,Qin:2014dqa}
\begin{align}M{\left(\tilde{\omega}_{n}^{2}, \vec{p}^{2}\right)}=\frac{B\left(\tilde{\omega}_{n}^{2}, \vec{p}\right)}{C\left(\tilde{\omega}_{n}^{2}, \vec{p}\right)}\end{align}
and the Euclidean constituent mass  $M^E := \{p^2 | p^2 > 0, p^2 = M{\left(\tilde{\omega}_{0}^{2}, \vec{p}^{2}\right)}\}$, which can be seen as a  free particle with mass $M^E$.\par
For finite size system, Eq. (\ref{DSE}) should be modified. Three momenta will be discretized by the boundary condition.
\begin{align}
\int \frac{d^{3} \vec{p}}{(2 \pi)^{3}} \rightarrow \frac{1}{V} \sum_{p_{k}}
\label{dsev}.
\end{align}
The allowed values of momentum modes depend on the selection of boundary conditions. For anti-periodic boundary condition (APBC), we have
\begin{align}
\vec{p}_{k}=\sum_{k_{i}=\pm 1,\pm 3, \ldots} \frac{k_{i} \pi }{L} \hat{e}_{i}
\end{align}
with  $L$ is the size of cubic box. Another boundary condition is multiple reflection expansion (MRE),   which introduces a IR cutoff in the momentum space and modifies the density of states \cite{Zhao2019}. \par
However, APBC  works on a cubic box and MRE becomes invalid for very small volume. In this work, we use MIT boundary condition. Under spherical MIT boundary condition, the allowed momentum values are given by the following eigen-equations
\begin{align}
j_{l_{\kappa}}(p R)=-\operatorname{sgn}(\kappa) \frac{p}{E+M} j_{\bar{l}_{\kappa}}(p R)
\end{align}
where
\begin{align*}
l_{\kappa}=\left\{\begin{array}{cl}-\kappa-1 & \text { for } \kappa<0 \\ \kappa & \text { for } \kappa>0\end{array}\right.
\end{align*}
\begin{align*}
\bar{l}_{\kappa}=\left\{\begin{array}{cl}-\kappa & \text { for } \kappa<0 \\ \kappa-1 & \text { for } \kappa>0\end{array}\right.
\end{align*}
$\kappa=\pm 1,\pm 2, \ldots$ and $j_{l}(x)$ is the $l$-th ordered spherical Bessel function. $R$ is the radius of sphere, $p$ is the allowed momentum value. The $M$ is the Euclidian constituent quark mass $M^E$.
\section{Finite volume effects on the chiral phase transition }
\label{QC}
Solving the DSEs at finite temperature and finite size, we obtain the numerical results of fully dressed quark propagator. We then study the chiral phase transition temperature $T_c$. The corresponding order parameter is
the chiral condensate. In the chiral limit, we have 
\begin{align}
-\langle\bar{\psi} \psi\rangle^{0}=N_{c} T \sum_{n=-\infty}^{\infty} \operatorname{tr}_{D} \int \frac{d^{3} p}{(2 \pi)^{3}} S\left( \vec{p},\tilde{\omega}_{n}\right)
\label{qc}
\end{align}
When $m \neq 0 $, Eq. (\ref{qc}) diverges, we do not have a well-defined chiral condensate. Hence, we employ the renormalized chiral condensate defined as \cite{Zong:2002ve,Maris:1997tm}
\begin{align}
-\langle\bar{\psi} \psi\rangle=N_{c} T \sum_{n=-\infty}^{\infty} \operatorname{tr}_{D} \int \frac{d^{3} p}{(2 \pi)^{3}} \left[S\left(\vec{p}, \tilde{\omega}_{n}\right)-S_0\left(\vec{p}, \tilde{\omega}_{n}\right)\right],
\end{align}
For finite size system it becomes 
\begin{align}
-\langle\bar{\psi} \psi\rangle_{V}=N_{c}\frac{T}{V} \sum_{k,n} \operatorname{tr_D}\left[S\left(\vec{p}_{k}, \omega_{n} \right)-S_{0}\left(\vec{p}_{k}, \omega_{n} \right)\right]
\label{qcv}
\end{align}
with $S_0\left(\vec{p}, \tilde{\omega}_{n}\right)$ being free quark  propagator and $S\left(\vec{p}_{k}, \omega_{n} \right)$ being the fully dressed quark propagator.\par
A direct computation of finite volume DSEs, i.e., Eqs. (\ref{DSE}, \ref{dsev}) is quite difficult.  The reason is obvious: In general, for the simplified Maris-Tandy model, the ultraviolet cutoff of the three momentum integral is $O(10)$ GeV and the summation of Matsubara frequency should be consistent with it.  As the volume increases and the temperature decreases, the number of allowed values of three momentum and Matsubara frequency is so large that it is difficult to calculate numerically. Some people approximates the momentum modes summation using an integral with an infrared cutoff \cite{Li2019,Bhattacharyya2013}. However, this approximation is known to get worse as the system size decreases.  In this work we adopt a new approximation, i.e., we rewrite Eq. (\ref{dsev}) as
\begin{align}
\int \frac{d^{3} \vec{p}}{(2 \pi)^{3}} \rightarrow \frac{1}{V} \sum_{|p_{k}|<\lambda}+\int_{|p_k|>\lambda} \frac{d^{3} \vec{p}}{(2 \pi)^{3}} 
\end{align}
As the high momentum modes are generally denser than low momentum modes, we therefore approximate the summation by an integral for high momentum modes.
Here $\lambda$ is an adjustable parameter, depending on volume and temperature. We can keep increasing it until the numerical result is stable. \par
For the summation of Matsubara frequency, symmetry of quark propagator can help us reduce the computation effort, 
\begin{align}
\mathcal{F}\left(\tilde{\omega}_{n},\vec{p}_k; T\right)=\mathcal{F}^{*}\left(\tilde{\omega}_{-n}, \vec{p}_k; T\right)
\end{align}
with $\mathcal{F}=A, B$ or $C$. At low temperature, the number of Matsubara frequency is of \ $\sim O(10^2)$.  For this reason, the numerical calculation is very difficult within DSEs framework when $T<0.1$ GeV. However, when $n$ is large enough, $\mathcal{F}$ is very smooth, so in the iteration we can reduce computing complexity by interpolation. The technique goes as follows: We take all the low Matsubara frequencies and keep only a few high  frequencies by means of a mapping
\begin{align}
n^\prime = \text{Int}[n^\gamma \cdot a + n\cdot(1-a)], 
\end{align}
where $n=1,2,3,...N$; $a=\frac{-N^\prime+N}{-N^\prime+(N^\prime)^\gamma}$ and $N$($N^\prime$) is the number of elements in array $n$($n^\prime$).  By controlling the scaling factor $\gamma$, we can map the evenly distributed array $n$ to an array $n^\prime$ that gets sparse as $n^\prime$ enlarges. We then solve the scalar functions $\mathcal{F}$ at frequencies $n'$, while at the rest frequencies the $\mathcal{F}$  are obtained by the cubic spline interpolation.
\par
When the chiral condensate is obtained, we can further study the chiral susceptibility \cite{Cui2017,Cui2015,Chakraborty2003,Smilga1996}
\begin{align}
\chi_{V}^{m}(T)=-\frac{\partial}{\partial m}\langle\bar{\psi} \psi\rangle_{V}.
\end{align}
In the chiral limit, chiral symmetry is restored via a second-order transition  at $T_c$ in which chiral susceptibility diverges. At nonzero current
mass, the chiral symmetry restoration
transition is replaced by a crossover. The pseudocritical
temperature $T_{pc}$ is obtained as the maxima of the chiral susceptibility with respect to temperature. \par
\begin{figure}
	\centering
	\includegraphics[width=0.4\textwidth]{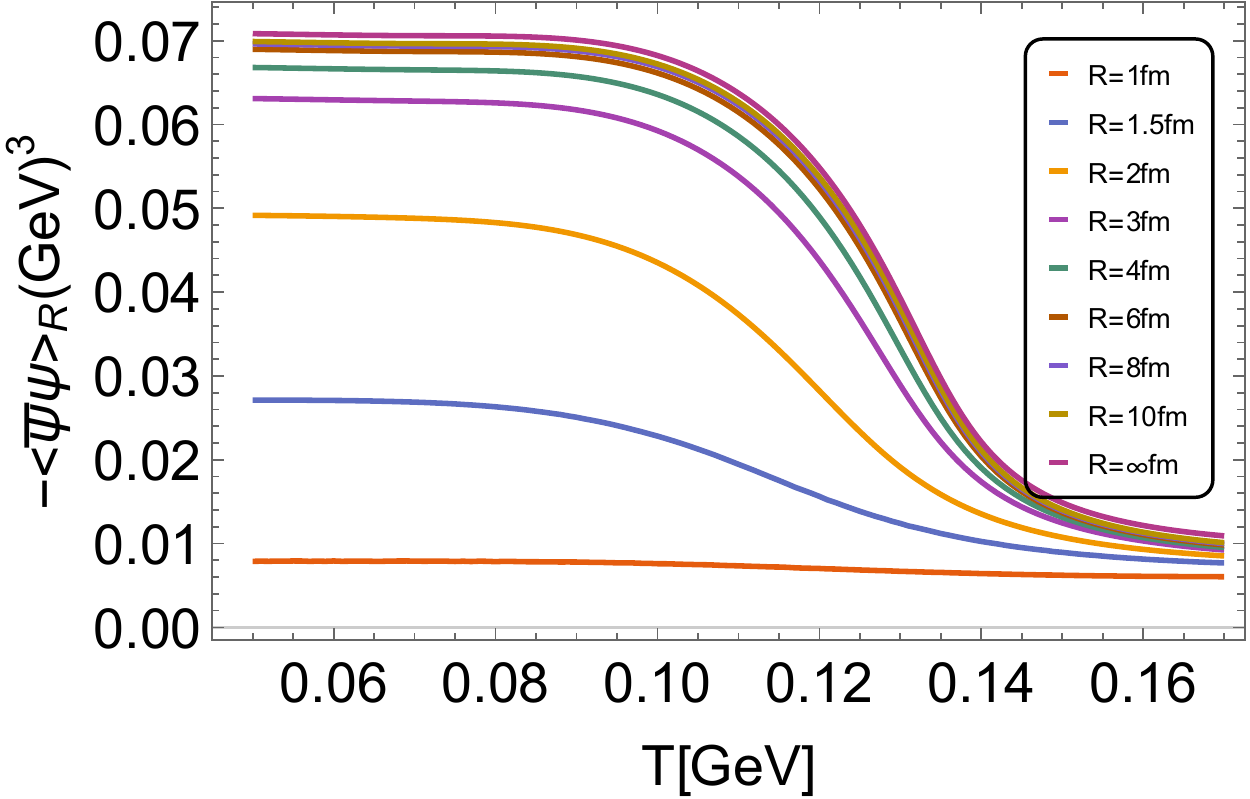}
	\includegraphics[width=0.4\textwidth]{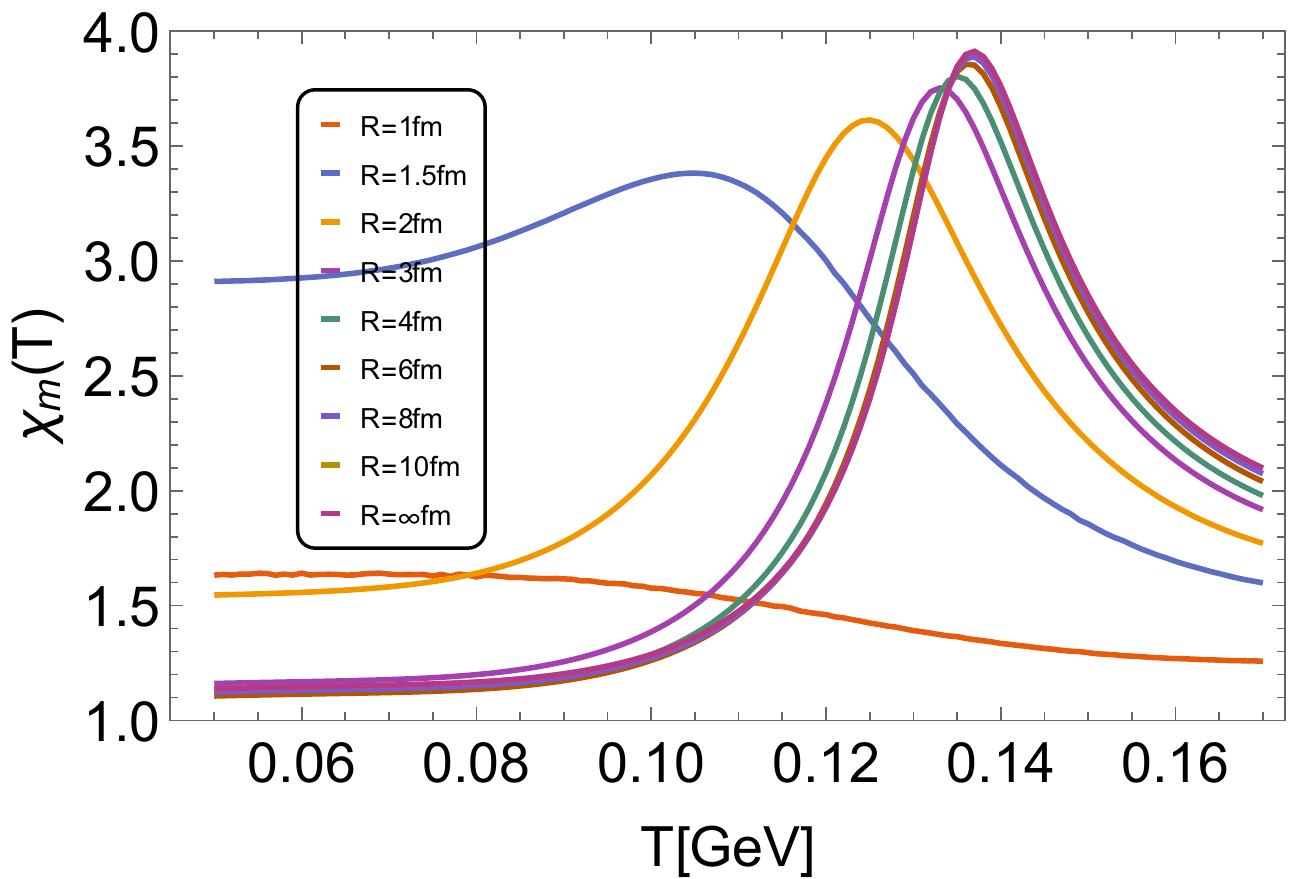}
	\caption{Upper panel: chiral condensate at different volumes. Lower panel: chiral susceptibility  at different volumes. For a spherical system,  when $R>10$ fm, the size of the system can be regarded as infinite. The volume change of the system starts to have a significant effect  when the $R<4$ fm and there is no phase transition in the system when $R<1.5$ fm.  }
	\label{cs}
\end{figure}
\begin{figure}
	\centering
	\includegraphics[width=0.4\textwidth]{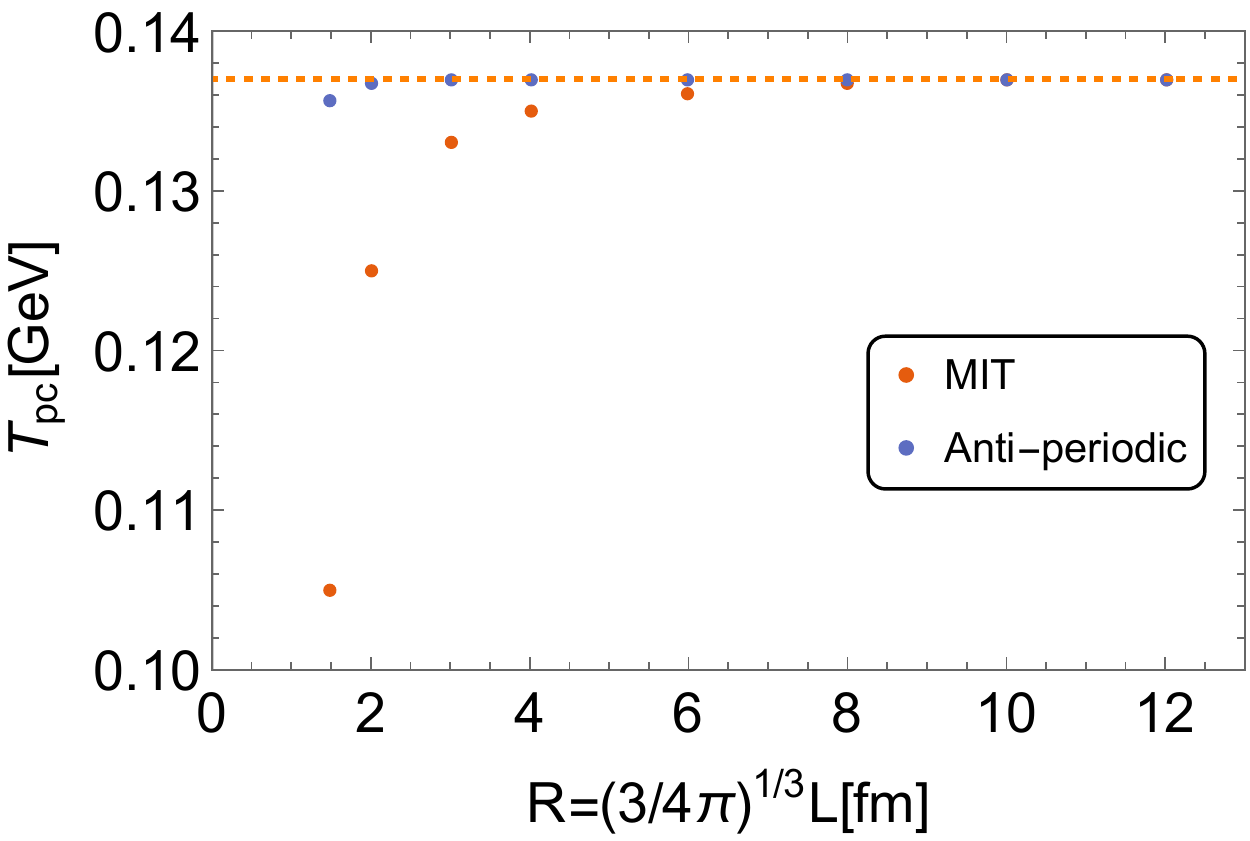}
	\caption{This figure shows the effect of the system shape on the $T_{pc}$. For a spherical system, the chiral crossover transition is more sensitive to the size of the system.}
	\label{cl}
\end{figure}
In this work, we choose the current-quark mass $m=4.8$ MeV, which is obtained by fitting the pion mass  ($m_\pi = 0.135$ GeV) and decay constant ($f_\pi= 0.095$ GeV) with the Bethe-Salpeter equation. When the volume of the system is infinite,  we have $T_{pc} =137$ MeV, close to the recent lattice QCD simulation value $T_{pc} =156.5 \pm 1.5$ MeV \cite{Bazavov2019a}. The numerical results show that  $T_{pc}$ gradually decreases and the curve of chiral susceptibility $\chi_{V}^{m}(T)$ becomes flat while the system size dwindling (see Fig. \ref{cs}). \par
When the volume is small enough, the influence of the shape of the system on $T_c$ cannot be ignored. Our earlier study shows that in a cubic box, when $L$ is greater than $3$ fm ($V\sim27\ \text{fm}^3$), the size of the system can be regarded as infinite \cite{Zhang2020,Shi2018,Shi2018a,Li2019}.  As Fischer at al pointed out in Ref. \cite{Luecker2010},  at zero temperature, the finite size will have large effects when cubic side length goes below\ $L=1.8\  \mathrm{fm}$ and chiral symmetry will get restored at small volumes with anti-periodic boundary condition.  However,  this work shows that, in contrast to a cubic system with APBC, the QGP phase transition in a spherical system using MIT condition is more sensitive to volume change (see Fig. \ref{cl}). When the radius is less than $10$ fm ($V\sim4000\ \text{fm}^3$),  the finite volume effect of the system emerges. The size change of the system starts to have a significant effect on $T_{pc}$ when the radius is decreased to $4$ fm ($V\sim270\ \text{fm}^3$). We argue that there is no chiral crossover transition in the system when the radius is less than 1.5 fm ($V\sim 14\ \text{fm}^3$). In other words, it is meaningless to discuss the chiral crossover  transition in a very small space size. Consider that DCSB can only occur in infinitely large systems in principle, chiral symmetry has restoratd when the size is small enough.\par
We note that similar results were found in Ref. \cite{Zhang2020}. Therein the NJL model study with MIT boundary condition shows that a system whose radius is above 14 fm can be regarded as an infinitely large system, while in a cubic the marginal size is $L=3$ fm. Hence, our results are closer to Ref. \cite{Zhang2020} than to DSEs with APBC \cite{Shi2018,Shi2018a,Li2019}. This indicates that the boundary condition potentially plays an important role when we want to simulate the realistic fireball produced in HICs.

\section{Summary and perspective}
\label{end}
Based on the DSEs formalism, we consider the influence of the finite volume on the chiral transition of QCD at finite temperature in a cubic and in a spherical. For the cubic volume, we use the widely adopted anti-periodic boundary condition and for the spherical volume we choose MIT boundary condition which had been used in NJL model \cite{Zhang2020}. While a cubic system within APBC could be regarded as an infinite system if the volume was greater than $27 \ \text{fm}^3$ ($L\sim3\ \text{fm}$), this calculation shows that the finite volume effects doesn't vanish until the sphere radius gets larger than 10 fm. Since the QGP system generated by the collision is closer to a sphere, we believe that the MIT boundary condition is closer to the real physics than the anti-periodic boundary condition. Our calculation further suggests that when the volume of the fireball produced in collision is less than  $270 \ \text{fm}^3$ ($R\sim 4\ \text{fm}$),  the volume effect becomes significant. And for $R<1.5$ fm, the chiral symmetry of system has restorated. This result also verifies Weinberg’s view: symmetry breaking can only occur in systems with a certain large size \cite{Weinberg:1996kr}. The study of QGP chiral transition is under way at the Relativistic Heavy Ion Collider (RHIC) and is planned at several future facilities, we expect that our results will be useful for recent and future experiments \cite{Loizides:2020tey,Senger2020,Harrison2003,Aggarwal:2010cw,Dainese:2016gch}. 
\section*{ACKNOWLEDGEMENTS}
We are grateful for insightful comments and suggestions from Zheng Zhang. This work is supported in part by the National Natural Science Foundation of China (under Grants No. 12075117, No. 11775112, No. 11535005, No. 11690030, and No. 11905104), and the Jiangsu Provincial Natural Science Foundation of China under Grant No. BK20180323.


\bibliography{temp_reference}

\begin{thebibliography}{52}%
\makeatletter
\providecommand \@ifxundefined [1]{%
 \@ifx{#1\undefined}
}%
\providecommand \@ifnum [1]{%
 \ifnum #1\expandafter \@firstoftwo
 \else \expandafter \@secondoftwo
 \fi
}%
\providecommand \@ifx [1]{%
 \ifx #1\expandafter \@firstoftwo
 \else \expandafter \@secondoftwo
 \fi
}%
\providecommand \natexlab [1]{#1}%
\providecommand \enquote  [1]{``#1''}%
\providecommand \bibnamefont  [1]{#1}%
\providecommand \bibfnamefont [1]{#1}%
\providecommand \citenamefont [1]{#1}%
\providecommand \href@noop [0]{\@secondoftwo}%
\providecommand \href [0]{\begingroup \@sanitize@url \@href}%
\providecommand \@href[1]{\@@startlink{#1}\@@href}%
\providecommand \@@href[1]{\endgroup#1\@@endlink}%
\providecommand \@sanitize@url [0]{\catcode `\\12\catcode `\$12\catcode
  `\&12\catcode `\#12\catcode `\^12\catcode `\_12\catcode `\%12\relax}%
\providecommand \@@startlink[1]{}%
\providecommand \@@endlink[0]{}%
\providecommand \url  [0]{\begingroup\@sanitize@url \@url }%
\providecommand \@url [1]{\endgroup\@href {#1}{\urlprefix }}%
\providecommand \urlprefix  [0]{URL }%
\providecommand \Eprint [0]{\href }%
\providecommand \doibase [0]{http://dx.doi.org/}%
\providecommand \selectlanguage [0]{\@gobble}%
\providecommand \bibinfo  [0]{\@secondoftwo}%
\providecommand \bibfield  [0]{\@secondoftwo}%
\providecommand \translation [1]{[#1]}%
\providecommand \BibitemOpen [0]{}%
\providecommand \bibitemStop [0]{}%
\providecommand \bibitemNoStop [0]{.\EOS\space}%
\providecommand \EOS [0]{\spacefactor3000\relax}%
\providecommand \BibitemShut  [1]{\csname bibitem#1\endcsname}%
\let\auto@bib@innerbib\@empty
\bibitem [{\citenamefont {Adams}\ \emph {et~al.}(2005)\citenamefont {Adams}
  \emph {et~al.}}]{Adams2005}%
  \BibitemOpen
  \bibfield  {author} {\bibinfo {author} {\bibfnamefont {J.}~\bibnamefont
  {Adams}} \emph {et~al.},\ }\href {\doibase 10.1016/j.nuclphysa.2005.03.085}
  {\bibfield  {journal} {\bibinfo  {journal} {Nuclear Physics A}\ }\textbf
  {\bibinfo {volume} {757}},\ \bibinfo {pages} {102} (\bibinfo {year}
  {2005})}\BibitemShut {NoStop}%
\bibitem [{\citenamefont {Shuryak}(2009)}]{Shuryak2009}%
  \BibitemOpen
  \bibfield  {author} {\bibinfo {author} {\bibfnamefont {E.}~\bibnamefont
  {Shuryak}},\ }\href {\doibase 10.1016/j.ppnp.2008.09.001} {\bibfield
  {journal} {\bibinfo  {journal} {Progress in Particle and Nuclear Physics}\
  }\textbf {\bibinfo {volume} {62}},\ \bibinfo {pages} {48} (\bibinfo {year}
  {2009})}\BibitemShut {NoStop}%
\bibitem [{\citenamefont {Bass}(1998)}]{Bass1998}%
  \BibitemOpen
  \bibfield  {author} {\bibinfo {author} {\bibfnamefont {S.}~\bibnamefont
  {Bass}},\ }\href {\doibase 10.1016/s0146-6410(98)00058-1} {\bibfield
  {journal} {\bibinfo  {journal} {Progress in Particle and Nuclear Physics}\
  }\textbf {\bibinfo {volume} {41}},\ \bibinfo {pages} {255} (\bibinfo {year}
  {1998})}\BibitemShut {NoStop}%
\bibitem [{\citenamefont {Gräf}\ \emph {et~al.}(2012)\citenamefont {Gräf},
  \citenamefont {Bleicher},\ and\ \citenamefont {Li}}]{Graef2012}%
  \BibitemOpen
  \bibfield  {author} {\bibinfo {author} {\bibfnamefont {G.}~\bibnamefont
  {Gräf}}, \bibinfo {author} {\bibfnamefont {M.}~\bibnamefont {Bleicher}}, \
  and\ \bibinfo {author} {\bibfnamefont {Q.}~\bibnamefont {Li}},\ }\href
  {\doibase 10.1103/physrevc.85.044901} {\bibfield  {journal} {\bibinfo
  {journal} {Physical Review C}\ }\textbf {\bibinfo {volume} {85}} (\bibinfo
  {year} {2012}),\ 10.1103/physrevc.85.044901}\BibitemShut {NoStop}%
\bibitem [{\citenamefont {Wang}\ \emph {et~al.}(2018)\citenamefont {Wang},
  \citenamefont {Xia},\ and\ \citenamefont {Zong}}]{Wang2018}%
  \BibitemOpen
  \bibfield  {author} {\bibinfo {author} {\bibfnamefont {Q.-W.}\ \bibnamefont
  {Wang}}, \bibinfo {author} {\bibfnamefont {Y.}~\bibnamefont {Xia}}, \ and\
  \bibinfo {author} {\bibfnamefont {H.-S.}\ \bibnamefont {Zong}},\ }\href
  {\doibase 10.1142/s0217732318502322} {\bibfield  {journal} {\bibinfo
  {journal} {Modern Physics Letters A}\ }\textbf {\bibinfo {volume} {33}},\
  \bibinfo {pages} {1850232} (\bibinfo {year} {2018})}\BibitemShut {NoStop}%
\bibitem [{\citenamefont {Abreu}\ \emph {et~al.}(2019)\citenamefont {Abreu},
  \citenamefont {Corr{\^{e}}a}, \citenamefont {Linhares},\ and\ \citenamefont
  {Malbouisson}}]{Abreu2019}%
  \BibitemOpen
  \bibfield  {author} {\bibinfo {author} {\bibfnamefont {L.~M.}\ \bibnamefont
  {Abreu}}, \bibinfo {author} {\bibfnamefont {E.~B.}\ \bibnamefont
  {Corr{\^{e}}a}}, \bibinfo {author} {\bibfnamefont {C.~A.}\ \bibnamefont
  {Linhares}}, \ and\ \bibinfo {author} {\bibfnamefont {A.~P.}\ \bibnamefont
  {Malbouisson}},\ }\href {\doibase 10.1103/physrevd.99.076001} {\bibfield
  {journal} {\bibinfo  {journal} {Physical Review D}\ }\textbf {\bibinfo
  {volume} {99}} (\bibinfo {year} {2019}),\
  10.1103/physrevd.99.076001}\BibitemShut {NoStop}%
\bibitem [{\citenamefont {Xia}\ \emph {et~al.}(2019)\citenamefont {Xia},
  \citenamefont {Wang}, \citenamefont {Feng},\ and\ \citenamefont
  {Zong}}]{xia2019}%
  \BibitemOpen
  \bibfield  {author} {\bibinfo {author} {\bibfnamefont {Y.}~\bibnamefont
  {Xia}}, \bibinfo {author} {\bibfnamefont {Q.}~\bibnamefont {Wang}}, \bibinfo
  {author} {\bibfnamefont {H.}~\bibnamefont {Feng}}, \ and\ \bibinfo {author}
  {\bibfnamefont {H.}~\bibnamefont {Zong}},\ }\href {\doibase
  10.1088/1674-1137/43/3/034101} {\bibfield  {journal} {\bibinfo  {journal}
  {Chinese Physics C}\ }\textbf {\bibinfo {volume} {43}},\ \bibinfo {pages}
  {034101} (\bibinfo {year} {2019})}\BibitemShut {NoStop}%
\bibitem [{\citenamefont {Bhattacharyya}\ \emph {et~al.}(2013)\citenamefont
  {Bhattacharyya}, \citenamefont {Deb}, \citenamefont {Ghosh}, \citenamefont
  {Ray},\ and\ \citenamefont {Sur}}]{Bhattacharyya2013}%
  \BibitemOpen
  \bibfield  {author} {\bibinfo {author} {\bibfnamefont {A.}~\bibnamefont
  {Bhattacharyya}}, \bibinfo {author} {\bibfnamefont {P.}~\bibnamefont {Deb}},
  \bibinfo {author} {\bibfnamefont {S.~K.}\ \bibnamefont {Ghosh}}, \bibinfo
  {author} {\bibfnamefont {R.}~\bibnamefont {Ray}}, \ and\ \bibinfo {author}
  {\bibfnamefont {S.}~\bibnamefont {Sur}},\ }\href {\doibase
  10.1103/physrevd.87.054009} {\bibfield  {journal} {\bibinfo  {journal}
  {Physical Review D}\ }\textbf {\bibinfo {volume} {87}} (\bibinfo {year}
  {2013}),\ 10.1103/physrevd.87.054009}\BibitemShut {NoStop}%
\bibitem [{\citenamefont {Pan}\ \emph {et~al.}(2017)\citenamefont {Pan},
  \citenamefont {Cui}, \citenamefont {Chang},\ and\ \citenamefont
  {Zong}}]{Pan2017}%
  \BibitemOpen
  \bibfield  {author} {\bibinfo {author} {\bibfnamefont {Z.}~\bibnamefont
  {Pan}}, \bibinfo {author} {\bibfnamefont {Z.-F.}\ \bibnamefont {Cui}},
  \bibinfo {author} {\bibfnamefont {C.-H.}\ \bibnamefont {Chang}}, \ and\
  \bibinfo {author} {\bibfnamefont {H.-S.}\ \bibnamefont {Zong}},\ }\href
  {\doibase 10.1142/s0217751x17500671} {\bibfield  {journal} {\bibinfo
  {journal} {International Journal of Modern Physics A}\ }\textbf {\bibinfo
  {volume} {32}},\ \bibinfo {pages} {1750067} (\bibinfo {year}
  {2017})}\BibitemShut {NoStop}%
\bibitem [{\citenamefont {Liu}\ \emph {et~al.}(2020)\citenamefont {Liu},
  \citenamefont {Lai}, \citenamefont {Shi},\ and\ \citenamefont
  {Zong}}]{Liu2020}%
  \BibitemOpen
  \bibfield  {author} {\bibinfo {author} {\bibfnamefont {R.-L.}\ \bibnamefont
  {Liu}}, \bibinfo {author} {\bibfnamefont {M.-Y.}\ \bibnamefont {Lai}},
  \bibinfo {author} {\bibfnamefont {C.}~\bibnamefont {Shi}}, \ and\ \bibinfo
  {author} {\bibfnamefont {H.-S.}\ \bibnamefont {Zong}},\ }\href {\doibase
  10.1103/physrevd.102.014014} {\bibfield  {journal} {\bibinfo  {journal}
  {Physical Review D}\ }\textbf {\bibinfo {volume} {102}} (\bibinfo {year}
  {2020}),\ 10.1103/physrevd.102.014014}\BibitemShut {NoStop}%
\bibitem [{\citenamefont {Tripolt}\ \emph {et~al.}(2014)\citenamefont
  {Tripolt}, \citenamefont {Braun}, \citenamefont {Klein},\ and\ \citenamefont
  {Schaefer}}]{Tripolt2014}%
  \BibitemOpen
  \bibfield  {author} {\bibinfo {author} {\bibfnamefont {R.-A.}\ \bibnamefont
  {Tripolt}}, \bibinfo {author} {\bibfnamefont {J.}~\bibnamefont {Braun}},
  \bibinfo {author} {\bibfnamefont {B.}~\bibnamefont {Klein}}, \ and\ \bibinfo
  {author} {\bibfnamefont {B.-J.}\ \bibnamefont {Schaefer}},\ }\href {\doibase
  10.1103/physrevd.90.054012} {\bibfield  {journal} {\bibinfo  {journal}
  {Physical Review D}\ }\textbf {\bibinfo {volume} {90}} (\bibinfo {year}
  {2014}),\ 10.1103/physrevd.90.054012}\BibitemShut {NoStop}%
\bibitem [{\citenamefont {Klein}(2017)}]{Klein2017}%
  \BibitemOpen
  \bibfield  {author} {\bibinfo {author} {\bibfnamefont {B.}~\bibnamefont
  {Klein}},\ }\href {\doibase 10.1016/j.physrep.2017.09.002} {\bibfield
  {journal} {\bibinfo  {journal} {Physics Reports}\ }\textbf {\bibinfo {volume}
  {707-708}},\ \bibinfo {pages} {1} (\bibinfo {year} {2017})}\BibitemShut
  {NoStop}%
\bibitem [{\citenamefont {Magdy}(2019)}]{Magdy2019}%
  \BibitemOpen
  \bibfield  {author} {\bibinfo {author} {\bibfnamefont {N.}~\bibnamefont
  {Magdy}},\ }\href {\doibase 10.3390/universe5040094} {\bibfield  {journal}
  {\bibinfo  {journal} {Universe}\ }\textbf {\bibinfo {volume} {5}},\ \bibinfo
  {pages} {94} (\bibinfo {year} {2019})}\BibitemShut {NoStop}%
\bibitem [{\citenamefont {Luecker}\ \emph {et~al.}(2010)\citenamefont
  {Luecker}, \citenamefont {Fischer},\ and\ \citenamefont
  {Williams}}]{Luecker2010}%
  \BibitemOpen
  \bibfield  {author} {\bibinfo {author} {\bibfnamefont {J.}~\bibnamefont
  {Luecker}}, \bibinfo {author} {\bibfnamefont {C.~S.}\ \bibnamefont
  {Fischer}}, \ and\ \bibinfo {author} {\bibfnamefont {R.}~\bibnamefont
  {Williams}},\ }\href {\doibase 10.1103/physrevd.81.094005} {\bibfield
  {journal} {\bibinfo  {journal} {Physical Review D}\ }\textbf {\bibinfo
  {volume} {81}} (\bibinfo {year} {2010}),\
  10.1103/physrevd.81.094005}\BibitemShut {NoStop}%
\bibitem [{\citenamefont {Li}\ \emph {et~al.}(2019)\citenamefont {Li},
  \citenamefont {Cui}, \citenamefont {Zhou}, \citenamefont {An}, \citenamefont
  {Zhang},\ and\ \citenamefont {Zong}}]{Li2019}%
  \BibitemOpen
  \bibfield  {author} {\bibinfo {author} {\bibfnamefont {B.-L.}\ \bibnamefont
  {Li}}, \bibinfo {author} {\bibfnamefont {Z.-F.}\ \bibnamefont {Cui}},
  \bibinfo {author} {\bibfnamefont {B.-W.}\ \bibnamefont {Zhou}}, \bibinfo
  {author} {\bibfnamefont {S.}~\bibnamefont {An}}, \bibinfo {author}
  {\bibfnamefont {L.-P.}\ \bibnamefont {Zhang}}, \ and\ \bibinfo {author}
  {\bibfnamefont {H.-S.}\ \bibnamefont {Zong}},\ }\href {\doibase
  10.1016/j.nuclphysb.2018.11.015} {\bibfield  {journal} {\bibinfo  {journal}
  {Nuclear Physics B}\ }\textbf {\bibinfo {volume} {938}},\ \bibinfo {pages}
  {298} (\bibinfo {year} {2019})}\BibitemShut {NoStop}%
\bibitem [{\citenamefont {Shi}\ \emph {et~al.}(2018{\natexlab{a}})\citenamefont
  {Shi}, \citenamefont {Xia}, \citenamefont {Jia},\ and\ \citenamefont
  {Zong}}]{Shi2018}%
  \BibitemOpen
  \bibfield  {author} {\bibinfo {author} {\bibfnamefont {C.}~\bibnamefont
  {Shi}}, \bibinfo {author} {\bibfnamefont {Y.}~\bibnamefont {Xia}}, \bibinfo
  {author} {\bibfnamefont {W.}~\bibnamefont {Jia}}, \ and\ \bibinfo {author}
  {\bibfnamefont {H.}~\bibnamefont {Zong}},\ }\href {\doibase
  10.1007/s11433-017-9177-4} {\bibfield  {journal} {\bibinfo  {journal}
  {Science China Physics, Mechanics {\&} Astronomy}\ }\textbf {\bibinfo
  {volume} {61}} (\bibinfo {year} {2018}{\natexlab{a}}),\
  10.1007/s11433-017-9177-4}\BibitemShut {NoStop}%
\bibitem [{\citenamefont {Shi}\ \emph {et~al.}(2018{\natexlab{b}})\citenamefont
  {Shi}, \citenamefont {Jia}, \citenamefont {Sun}, \citenamefont {Zhang},\ and\
  \citenamefont {Zong}}]{Shi2018a}%
  \BibitemOpen
  \bibfield  {author} {\bibinfo {author} {\bibfnamefont {C.}~\bibnamefont
  {Shi}}, \bibinfo {author} {\bibfnamefont {W.}~\bibnamefont {Jia}}, \bibinfo
  {author} {\bibfnamefont {A.}~\bibnamefont {Sun}}, \bibinfo {author}
  {\bibfnamefont {L.}~\bibnamefont {Zhang}}, \ and\ \bibinfo {author}
  {\bibfnamefont {H.}~\bibnamefont {Zong}},\ }\href {\doibase
  10.1088/1674-1137/42/2/023101} {\bibfield  {journal} {\bibinfo  {journal}
  {Chinese Physics C}\ }\textbf {\bibinfo {volume} {42}},\ \bibinfo {pages}
  {023101} (\bibinfo {year} {2018}{\natexlab{b}})}\BibitemShut {NoStop}%
\bibitem [{\citenamefont {Shi}\ \emph {et~al.}(2020)\citenamefont {Shi},
  \citenamefont {He}, \citenamefont {Jia}, \citenamefont {Wang}, \citenamefont
  {Xu},\ and\ \citenamefont {Zong}}]{Shi2020}%
  \BibitemOpen
  \bibfield  {author} {\bibinfo {author} {\bibfnamefont {C.}~\bibnamefont
  {Shi}}, \bibinfo {author} {\bibfnamefont {X.-T.}\ \bibnamefont {He}},
  \bibinfo {author} {\bibfnamefont {W.-B.}\ \bibnamefont {Jia}}, \bibinfo
  {author} {\bibfnamefont {Q.-W.}\ \bibnamefont {Wang}}, \bibinfo {author}
  {\bibfnamefont {S.-S.}\ \bibnamefont {Xu}}, \ and\ \bibinfo {author}
  {\bibfnamefont {H.-S.}\ \bibnamefont {Zong}},\ }\href {\doibase
  10.1007/jhep06(2020)122} {\bibfield  {journal} {\bibinfo  {journal} {Journal
  of High Energy Physics}\ }\textbf {\bibinfo {volume} {2020}} (\bibinfo {year}
  {2020}),\ 10.1007/jhep06(2020)122}\BibitemShut {NoStop}%
\bibitem [{\citenamefont {Zhao}\ \emph {et~al.}(2019)\citenamefont {Zhao},
  \citenamefont {Zhang}, \citenamefont {Zhang},\ and\ \citenamefont
  {Zong}}]{Zhao2019}%
  \BibitemOpen
  \bibfield  {author} {\bibinfo {author} {\bibfnamefont {Y.-P.}\ \bibnamefont
  {Zhao}}, \bibinfo {author} {\bibfnamefont {R.-R.}\ \bibnamefont {Zhang}},
  \bibinfo {author} {\bibfnamefont {H.}~\bibnamefont {Zhang}}, \ and\ \bibinfo
  {author} {\bibfnamefont {H.-S.}\ \bibnamefont {Zong}},\ }\href {\doibase
  10.1088/1674-1137/43/6/063101} {\bibfield  {journal} {\bibinfo  {journal}
  {Chinese Physics C}\ }\textbf {\bibinfo {volume} {43}},\ \bibinfo {pages}
  {063101} (\bibinfo {year} {2019})}\BibitemShut {NoStop}%
\bibitem [{\citenamefont {Zhang}\ \emph {et~al.}(2020)\citenamefont {Zhang},
  \citenamefont {Shi},\ and\ \citenamefont {Zong}}]{Zhang2020}%
  \BibitemOpen
  \bibfield  {author} {\bibinfo {author} {\bibfnamefont {Z.}~\bibnamefont
  {Zhang}}, \bibinfo {author} {\bibfnamefont {C.}~\bibnamefont {Shi}}, \ and\
  \bibinfo {author} {\bibfnamefont {H.-S.}\ \bibnamefont {Zong}},\ }\href
  {\doibase 10.1103/physrevd.101.043006} {\bibfield  {journal} {\bibinfo
  {journal} {Physical Review D}\ }\textbf {\bibinfo {volume} {101}} (\bibinfo
  {year} {2020}),\ 10.1103/physrevd.101.043006}\BibitemShut {NoStop}%
\bibitem [{\citenamefont {Roberts}\ and\ \citenamefont
  {Schmidt}(2000)}]{ROBERTS2000S1}%
  \BibitemOpen
  \bibfield  {author} {\bibinfo {author} {\bibfnamefont {C.}~\bibnamefont
  {Roberts}}\ and\ \bibinfo {author} {\bibfnamefont {S.}~\bibnamefont
  {Schmidt}},\ }\href {\doibase https://doi.org/10.1016/S0146-6410(00)90011-5}
  {\bibfield  {journal} {\bibinfo  {journal} {Progress in Particle and Nuclear
  Physics}\ }\textbf {\bibinfo {volume} {45}},\ \bibinfo {pages} {S1 }
  (\bibinfo {year} {2000})}\BibitemShut {NoStop}%
\bibitem [{\citenamefont {Hayano}\ and\ \citenamefont
  {Hatsuda}(2010)}]{RevModPhys.82.2949}%
  \BibitemOpen
  \bibfield  {author} {\bibinfo {author} {\bibfnamefont {R.~S.}\ \bibnamefont
  {Hayano}}\ and\ \bibinfo {author} {\bibfnamefont {T.}~\bibnamefont
  {Hatsuda}},\ }\href {\doibase 10.1103/RevModPhys.82.2949} {\bibfield
  {journal} {\bibinfo  {journal} {Rev. Mod. Phys.}\ }\textbf {\bibinfo {volume}
  {82}},\ \bibinfo {pages} {2949} (\bibinfo {year} {2010})}\BibitemShut
  {NoStop}%
\bibitem [{\citenamefont {Maris}\ and\ \citenamefont
  {Roberts}(1997)}]{Maris:1997tm}%
  \BibitemOpen
  \bibfield  {author} {\bibinfo {author} {\bibfnamefont {P.}~\bibnamefont
  {Maris}}\ and\ \bibinfo {author} {\bibfnamefont {C.~D.}\ \bibnamefont
  {Roberts}},\ }\href {\doibase 10.1103/PhysRevC.56.3369} {\bibfield  {journal}
  {\bibinfo  {journal} {Phys. Rev. C}\ }\textbf {\bibinfo {volume} {56}},\
  \bibinfo {pages} {3369} (\bibinfo {year} {1997})},\ \Eprint
  {http://arxiv.org/abs/nucl-th/9708029} {arXiv:nucl-th/9708029} \BibitemShut
  {NoStop}%
\bibitem [{\citenamefont {Maris}\ and\ \citenamefont
  {Tandy}(1999)}]{Maris:1999nt}%
  \BibitemOpen
  \bibfield  {author} {\bibinfo {author} {\bibfnamefont {P.}~\bibnamefont
  {Maris}}\ and\ \bibinfo {author} {\bibfnamefont {P.~C.}\ \bibnamefont
  {Tandy}},\ }\href {\doibase 10.1103/PhysRevC.60.055214} {\bibfield  {journal}
  {\bibinfo  {journal} {Phys. Rev.}\ }\textbf {\bibinfo {volume} {C60}},\
  \bibinfo {pages} {055214} (\bibinfo {year} {1999})},\ \Eprint
  {http://arxiv.org/abs/nucl-th/9905056} {arXiv:nucl-th/9905056 [nucl-th]}
  \BibitemShut {NoStop}%
\bibitem [{\citenamefont {Fischer}\ \emph {et~al.}(2015)\citenamefont
  {Fischer}, \citenamefont {Kubrak},\ and\ \citenamefont
  {Williams}}]{Fischer:2014cfa}%
  \BibitemOpen
  \bibfield  {author} {\bibinfo {author} {\bibfnamefont {C.~S.}\ \bibnamefont
  {Fischer}}, \bibinfo {author} {\bibfnamefont {S.}~\bibnamefont {Kubrak}}, \
  and\ \bibinfo {author} {\bibfnamefont {R.}~\bibnamefont {Williams}},\ }\href
  {\doibase 10.1140/epja/i2015-15010-7} {\bibfield  {journal} {\bibinfo
  {journal} {Eur. Phys. J.}\ }\textbf {\bibinfo {volume} {A51}},\ \bibinfo
  {pages} {10} (\bibinfo {year} {2015})},\ \Eprint
  {http://arxiv.org/abs/1409.5076} {arXiv:1409.5076 [hep-ph]} \BibitemShut
  {NoStop}%
\bibitem [{\citenamefont {Hilger}\ \emph {et~al.}(2017)\citenamefont {Hilger},
  \citenamefont {Gomez-Rocha},\ and\ \citenamefont
  {Krassnigg}}]{Hilger:2015ora}%
  \BibitemOpen
  \bibfield  {author} {\bibinfo {author} {\bibfnamefont {T.}~\bibnamefont
  {Hilger}}, \bibinfo {author} {\bibfnamefont {M.}~\bibnamefont {Gomez-Rocha}},
  \ and\ \bibinfo {author} {\bibfnamefont {A.}~\bibnamefont {Krassnigg}},\
  }\href {\doibase 10.1140/epjc/s10052-017-5163-4} {\bibfield  {journal}
  {\bibinfo  {journal} {Eur. Phys. J.}\ }\textbf {\bibinfo {volume} {C77}},\
  \bibinfo {pages} {625} (\bibinfo {year} {2017})},\ \Eprint
  {http://arxiv.org/abs/1508.07183} {arXiv:1508.07183 [hep-ph]} \BibitemShut
  {NoStop}%
\bibitem [{\citenamefont {H\"oll}\ \emph {et~al.}(2005)\citenamefont {H\"oll},
  \citenamefont {Krassnigg}, \citenamefont {Maris}, \citenamefont {Roberts},\
  and\ \citenamefont {Wright}}]{PhysRevC.71.065204}%
  \BibitemOpen
  \bibfield  {author} {\bibinfo {author} {\bibfnamefont {A.}~\bibnamefont
  {H\"oll}}, \bibinfo {author} {\bibfnamefont {A.}~\bibnamefont {Krassnigg}},
  \bibinfo {author} {\bibfnamefont {P.}~\bibnamefont {Maris}}, \bibinfo
  {author} {\bibfnamefont {C.~D.}\ \bibnamefont {Roberts}}, \ and\ \bibinfo
  {author} {\bibfnamefont {S.~V.}\ \bibnamefont {Wright}},\ }\href {\doibase
  10.1103/PhysRevC.71.065204} {\bibfield  {journal} {\bibinfo  {journal} {Phys.
  Rev. C}\ }\textbf {\bibinfo {volume} {71}},\ \bibinfo {pages} {065204}
  (\bibinfo {year} {2005})}\BibitemShut {NoStop}%
\bibitem [{\citenamefont {Bhagwat}\ and\ \citenamefont
  {Maris}(2008)}]{PhysRevC.77.025203}%
  \BibitemOpen
  \bibfield  {author} {\bibinfo {author} {\bibfnamefont {M.~S.}\ \bibnamefont
  {Bhagwat}}\ and\ \bibinfo {author} {\bibfnamefont {P.}~\bibnamefont
  {Maris}},\ }\href {\doibase 10.1103/PhysRevC.77.025203} {\bibfield  {journal}
  {\bibinfo  {journal} {Phys. Rev. C}\ }\textbf {\bibinfo {volume} {77}},\
  \bibinfo {pages} {025203} (\bibinfo {year} {2008})}\BibitemShut {NoStop}%
\bibitem [{\citenamefont {Maris}\ and\ \citenamefont
  {Tandy}(2000)}]{PhysRevC.61.045202}%
  \BibitemOpen
  \bibfield  {author} {\bibinfo {author} {\bibfnamefont {P.}~\bibnamefont
  {Maris}}\ and\ \bibinfo {author} {\bibfnamefont {P.~C.}\ \bibnamefont
  {Tandy}},\ }\href {\doibase 10.1103/PhysRevC.61.045202} {\bibfield  {journal}
  {\bibinfo  {journal} {Phys. Rev. C}\ }\textbf {\bibinfo {volume} {61}},\
  \bibinfo {pages} {045202} (\bibinfo {year} {2000})}\BibitemShut {NoStop}%
\bibitem [{\citenamefont {Roberts}(2008)}]{Roberts2008}%
  \BibitemOpen
  \bibfield  {author} {\bibinfo {author} {\bibfnamefont {C.}~\bibnamefont
  {Roberts}},\ }\href {\doibase 10.1016/j.ppnp.2007.12.034} {\bibfield
  {journal} {\bibinfo  {journal} {Progress in Particle and Nuclear Physics}\
  }\textbf {\bibinfo {volume} {61}},\ \bibinfo {pages} {50} (\bibinfo {year}
  {2008})}\BibitemShut {NoStop}%
\bibitem [{\citenamefont {Xu}\ \emph {et~al.}(2019)\citenamefont {Xu},
  \citenamefont {Binosi}, \citenamefont {Cui}, \citenamefont {Li},
  \citenamefont {Roberts}, \citenamefont {Xu},\ and\ \citenamefont
  {Zong}}]{Xu2019}%
  \BibitemOpen
  \bibfield  {author} {\bibinfo {author} {\bibfnamefont {Y.-Z.}\ \bibnamefont
  {Xu}}, \bibinfo {author} {\bibfnamefont {D.}~\bibnamefont {Binosi}}, \bibinfo
  {author} {\bibfnamefont {Z.-F.}\ \bibnamefont {Cui}}, \bibinfo {author}
  {\bibfnamefont {B.-L.}\ \bibnamefont {Li}}, \bibinfo {author} {\bibfnamefont
  {C.}~\bibnamefont {Roberts}}, \bibinfo {author} {\bibfnamefont {S.-S.}\
  \bibnamefont {Xu}}, \ and\ \bibinfo {author} {\bibfnamefont {H.-S.}\
  \bibnamefont {Zong}},\ }\href {\doibase 10.1103/physrevd.100.114038}
  {\bibfield  {journal} {\bibinfo  {journal} {Physical Review D}\ }\textbf
  {\bibinfo {volume} {100}} (\bibinfo {year} {2019}),\
  10.1103/physrevd.100.114038}\BibitemShut {NoStop}%
\bibitem [{\citenamefont {Qin}\ \emph {et~al.}(2011)\citenamefont {Qin},
  \citenamefont {Chang}, \citenamefont {Chen}, \citenamefont {Liu},\ and\
  \citenamefont {Roberts}}]{Qin:2010nq}%
  \BibitemOpen
  \bibfield  {author} {\bibinfo {author} {\bibfnamefont {S.-x.}\ \bibnamefont
  {Qin}}, \bibinfo {author} {\bibfnamefont {L.}~\bibnamefont {Chang}}, \bibinfo
  {author} {\bibfnamefont {H.}~\bibnamefont {Chen}}, \bibinfo {author}
  {\bibfnamefont {Y.-x.}\ \bibnamefont {Liu}}, \ and\ \bibinfo {author}
  {\bibfnamefont {C.~D.}\ \bibnamefont {Roberts}},\ }\href {\doibase
  10.1103/PhysRevLett.106.172301} {\bibfield  {journal} {\bibinfo  {journal}
  {Phys. Rev. Lett.}\ }\textbf {\bibinfo {volume} {106}},\ \bibinfo {pages}
  {172301} (\bibinfo {year} {2011})},\ \Eprint {http://arxiv.org/abs/1011.2876}
  {arXiv:1011.2876 [nucl-th]} \BibitemShut {NoStop}%
\bibitem [{\citenamefont {Roberts}\ and\ \citenamefont
  {Williams}(1994)}]{Roberts1994}%
  \BibitemOpen
  \bibfield  {author} {\bibinfo {author} {\bibfnamefont {C.~D.}\ \bibnamefont
  {Roberts}}\ and\ \bibinfo {author} {\bibfnamefont {A.~G.}\ \bibnamefont
  {Williams}},\ }\href {\doibase 10.1016/0146-6410(94)90049-3} {\bibfield
  {journal} {\bibinfo  {journal} {Progress in Particle and Nuclear Physics}\
  }\textbf {\bibinfo {volume} {33}},\ \bibinfo {pages} {477} (\bibinfo {year}
  {1994})}\BibitemShut {NoStop}%
\bibitem [{\citenamefont {Roberts}\ \emph {et~al.}(2007)\citenamefont
  {Roberts}, \citenamefont {Bhagwat}, \citenamefont {Höll},\ and\
  \citenamefont {Wright}}]{Roberts2007}%
  \BibitemOpen
  \bibfield  {author} {\bibinfo {author} {\bibfnamefont {C.~D.}\ \bibnamefont
  {Roberts}}, \bibinfo {author} {\bibfnamefont {M.~S.}\ \bibnamefont
  {Bhagwat}}, \bibinfo {author} {\bibfnamefont {A.}~\bibnamefont {Höll}}, \
  and\ \bibinfo {author} {\bibfnamefont {S.~V.}\ \bibnamefont {Wright}},\
  }\href {\doibase 10.1140/epjst/e2007-00003-5} {\bibfield  {journal} {\bibinfo
   {journal} {The European Physical Journal Special Topics}\ }\textbf {\bibinfo
  {volume} {140}},\ \bibinfo {pages} {53} (\bibinfo {year} {2007})}\BibitemShut
  {NoStop}%
\bibitem [{\citenamefont {McLerran}\ and\ \citenamefont
  {Pisarski}(2007)}]{McLerran2007}%
  \BibitemOpen
  \bibfield  {author} {\bibinfo {author} {\bibfnamefont {L.}~\bibnamefont
  {McLerran}}\ and\ \bibinfo {author} {\bibfnamefont {R.~D.}\ \bibnamefont
  {Pisarski}},\ }\href {\doibase 10.1016/j.nuclphysa.2007.08.013} {\bibfield
  {journal} {\bibinfo  {journal} {Nuclear Physics A}\ }\textbf {\bibinfo
  {volume} {796}},\ \bibinfo {pages} {83} (\bibinfo {year} {2007})}\BibitemShut
  {NoStop}%
\bibitem [{\citenamefont {Bashir}\ \emph {et~al.}(2012)\citenamefont {Bashir},
  \citenamefont {Chang}, \citenamefont {Cloet}, \citenamefont {El-Bennich},
  \citenamefont {Liu}, \citenamefont {Roberts},\ and\ \citenamefont
  {Tandy}}]{Bashir:2012fs}%
  \BibitemOpen
  \bibfield  {author} {\bibinfo {author} {\bibfnamefont {A.}~\bibnamefont
  {Bashir}}, \bibinfo {author} {\bibfnamefont {L.}~\bibnamefont {Chang}},
  \bibinfo {author} {\bibfnamefont {I.~C.}\ \bibnamefont {Cloet}}, \bibinfo
  {author} {\bibfnamefont {B.}~\bibnamefont {El-Bennich}}, \bibinfo {author}
  {\bibfnamefont {Y.-X.}\ \bibnamefont {Liu}}, \bibinfo {author} {\bibfnamefont
  {C.~D.}\ \bibnamefont {Roberts}}, \ and\ \bibinfo {author} {\bibfnamefont
  {P.~C.}\ \bibnamefont {Tandy}},\ }\href {\doibase 10.1088/0253-6102/58/1/16}
  {\bibfield  {journal} {\bibinfo  {journal} {Commun. Theor. Phys.}\ }\textbf
  {\bibinfo {volume} {58}},\ \bibinfo {pages} {79} (\bibinfo {year} {2012})},\
  \Eprint {http://arxiv.org/abs/1201.3366} {arXiv:1201.3366 [nucl-th]}
  \BibitemShut {NoStop}%
\bibitem [{\citenamefont {Chang}\ \emph {et~al.}(2011)\citenamefont {Chang},
  \citenamefont {Roberts},\ and\ \citenamefont {Tandy}}]{Chang:2011vu}%
  \BibitemOpen
  \bibfield  {author} {\bibinfo {author} {\bibfnamefont {L.}~\bibnamefont
  {Chang}}, \bibinfo {author} {\bibfnamefont {C.~D.}\ \bibnamefont {Roberts}},
  \ and\ \bibinfo {author} {\bibfnamefont {P.~C.}\ \bibnamefont {Tandy}},\
  }\href@noop {} {\bibfield  {journal} {\bibinfo  {journal} {Chin. J. Phys.}\
  }\textbf {\bibinfo {volume} {49}},\ \bibinfo {pages} {955} (\bibinfo {year}
  {2011})},\ \Eprint {http://arxiv.org/abs/1107.4003} {arXiv:1107.4003
  [nucl-th]} \BibitemShut {NoStop}%
\bibitem [{\citenamefont {Maris}\ and\ \citenamefont
  {Roberts}(2003)}]{Maris:2003vk}%
  \BibitemOpen
  \bibfield  {author} {\bibinfo {author} {\bibfnamefont {P.}~\bibnamefont
  {Maris}}\ and\ \bibinfo {author} {\bibfnamefont {C.~D.}\ \bibnamefont
  {Roberts}},\ }\href {\doibase 10.1142/S0218301303001326} {\bibfield
  {journal} {\bibinfo  {journal} {Int. J. Mod. Phys.}\ }\textbf {\bibinfo
  {volume} {E12}},\ \bibinfo {pages} {297} (\bibinfo {year} {2003})},\ \Eprint
  {http://arxiv.org/abs/nucl-th/0301049} {arXiv:nucl-th/0301049 [nucl-th]}
  \BibitemShut {NoStop}%
\bibitem [{\citenamefont {M\"uller}(2011)}]{Muller:2011awa}%
  \BibitemOpen
  \bibfield  {author} {\bibinfo {author} {\bibfnamefont {J.~A.}\ \bibnamefont
  {M\"uller}},\ }\emph {\bibinfo {title} {{A Dyson-Schwinger Approach to Finite
  Temperature QCD}}},\ \href@noop {} {Ph.D. thesis},\ \bibinfo  {school}
  {Darmstadt, Tech. Hochsch.} (\bibinfo {year} {2011})\BibitemShut {NoStop}%
\bibitem [{\citenamefont {Qin}\ and\ \citenamefont
  {Rischke}(2014)}]{Qin:2014dqa}%
  \BibitemOpen
  \bibfield  {author} {\bibinfo {author} {\bibfnamefont {S.-x.}\ \bibnamefont
  {Qin}}\ and\ \bibinfo {author} {\bibfnamefont {D.~H.}\ \bibnamefont
  {Rischke}},\ }\href {\doibase 10.1016/j.physletb.2014.05.060} {\bibfield
  {journal} {\bibinfo  {journal} {Phys. Lett.}\ }\textbf {\bibinfo {volume}
  {B734}},\ \bibinfo {pages} {157} (\bibinfo {year} {2014})},\ \Eprint
  {http://arxiv.org/abs/1403.3025} {arXiv:1403.3025 [nucl-th]} \BibitemShut
  {NoStop}%
\bibitem [{\citenamefont {Zong}\ \emph {et~al.}(2003)\citenamefont {Zong},
  \citenamefont {Ping}, \citenamefont {Yang}, \citenamefont {Lu},\ and\
  \citenamefont {Wang}}]{Zong:2002ve}%
  \BibitemOpen
  \bibfield  {author} {\bibinfo {author} {\bibfnamefont {H.-s.}\ \bibnamefont
  {Zong}}, \bibinfo {author} {\bibfnamefont {J.-l.}\ \bibnamefont {Ping}},
  \bibinfo {author} {\bibfnamefont {H.-t.}\ \bibnamefont {Yang}}, \bibinfo
  {author} {\bibfnamefont {X.-f.}\ \bibnamefont {Lu}}, \ and\ \bibinfo {author}
  {\bibfnamefont {F.}~\bibnamefont {Wang}},\ }\href {\doibase
  10.1103/PhysRevD.67.074004} {\bibfield  {journal} {\bibinfo  {journal} {Phys.
  Rev. D}\ }\textbf {\bibinfo {volume} {67}},\ \bibinfo {pages} {074004}
  (\bibinfo {year} {2003})},\ \Eprint {http://arxiv.org/abs/nucl-th/0201001}
  {arXiv:nucl-th/0201001} \BibitemShut {NoStop}%
\bibitem [{\citenamefont {Cui}\ \emph {et~al.}(2017)\citenamefont {Cui},
  \citenamefont {Zhang},\ and\ \citenamefont {Zong}}]{Cui2017}%
  \BibitemOpen
  \bibfield  {author} {\bibinfo {author} {\bibfnamefont {Z.-F.}\ \bibnamefont
  {Cui}}, \bibinfo {author} {\bibfnamefont {J.-L.}\ \bibnamefont {Zhang}}, \
  and\ \bibinfo {author} {\bibfnamefont {H.-S.}\ \bibnamefont {Zong}},\ }\href
  {\doibase 10.1038/srep45937} {\bibfield  {journal} {\bibinfo  {journal}
  {Scientific Reports}\ }\textbf {\bibinfo {volume} {7}} (\bibinfo {year}
  {2017}),\ 10.1038/srep45937}\BibitemShut {NoStop}%
\bibitem [{\citenamefont {Cui}\ \emph {et~al.}(2015)\citenamefont {Cui},
  \citenamefont {Hou}, \citenamefont {Shi}, \citenamefont {Wang},\ and\
  \citenamefont {Zong}}]{Cui2015}%
  \BibitemOpen
  \bibfield  {author} {\bibinfo {author} {\bibfnamefont {Z.-F.}\ \bibnamefont
  {Cui}}, \bibinfo {author} {\bibfnamefont {F.-Y.}\ \bibnamefont {Hou}},
  \bibinfo {author} {\bibfnamefont {Y.-M.}\ \bibnamefont {Shi}}, \bibinfo
  {author} {\bibfnamefont {Y.-L.}\ \bibnamefont {Wang}}, \ and\ \bibinfo
  {author} {\bibfnamefont {H.-S.}\ \bibnamefont {Zong}},\ }\href {\doibase
  10.1016/j.aop.2015.03.025} {\bibfield  {journal} {\bibinfo  {journal} {Annals
  of Physics}\ }\textbf {\bibinfo {volume} {358}},\ \bibinfo {pages} {172}
  (\bibinfo {year} {2015})}\BibitemShut {NoStop}%
\bibitem [{\citenamefont {Chakraborty}\ \emph {et~al.}(2003)\citenamefont
  {Chakraborty}, \citenamefont {Mustafa},\ and\ \citenamefont
  {Thoma}}]{Chakraborty2003}%
  \BibitemOpen
  \bibfield  {author} {\bibinfo {author} {\bibfnamefont {P.}~\bibnamefont
  {Chakraborty}}, \bibinfo {author} {\bibfnamefont {M.~G.}\ \bibnamefont
  {Mustafa}}, \ and\ \bibinfo {author} {\bibfnamefont {M.~H.}\ \bibnamefont
  {Thoma}},\ }\href {\doibase 10.1103/physrevd.67.114004} {\bibfield  {journal}
  {\bibinfo  {journal} {Physical Review D}\ }\textbf {\bibinfo {volume} {67}}
  (\bibinfo {year} {2003}),\ 10.1103/physrevd.67.114004}\BibitemShut {NoStop}%
\bibitem [{\citenamefont {Smilga}\ and\ \citenamefont
  {Verbaarschot}(1996)}]{Smilga1996}%
  \BibitemOpen
  \bibfield  {author} {\bibinfo {author} {\bibfnamefont {A.}~\bibnamefont
  {Smilga}}\ and\ \bibinfo {author} {\bibfnamefont {J.~J.~M.}\ \bibnamefont
  {Verbaarschot}},\ }\href {\doibase 10.1103/physrevd.54.1087} {\bibfield
  {journal} {\bibinfo  {journal} {Physical Review D}\ }\textbf {\bibinfo
  {volume} {54}},\ \bibinfo {pages} {1087} (\bibinfo {year}
  {1996})}\BibitemShut {NoStop}%
\bibitem [{\citenamefont {Bazavov}\ \emph {et~al.}(2019)\citenamefont
  {Bazavov}, \citenamefont {Ding}, \citenamefont {Hegde}, \citenamefont
  {Kaczmarek}, \citenamefont {Karsch}, \citenamefont {Karthik}, \citenamefont
  {Laermann}, \citenamefont {Lahiri}, \citenamefont {Larsen}, \citenamefont
  {Li}, \citenamefont {Mukherjee}, \citenamefont {Ohno}, \citenamefont
  {Petreczky}, \citenamefont {Sandmeyer}, \citenamefont {Schmidt},
  \citenamefont {Sharma},\ and\ \citenamefont {Steinbrecher}}]{Bazavov2019a}%
  \BibitemOpen
  \bibfield  {author} {\bibinfo {author} {\bibfnamefont {A.}~\bibnamefont
  {Bazavov}}, \bibinfo {author} {\bibfnamefont {H.-T.}\ \bibnamefont {Ding}},
  \bibinfo {author} {\bibfnamefont {P.}~\bibnamefont {Hegde}}, \bibinfo
  {author} {\bibfnamefont {O.}~\bibnamefont {Kaczmarek}}, \bibinfo {author}
  {\bibfnamefont {F.}~\bibnamefont {Karsch}}, \bibinfo {author} {\bibfnamefont
  {N.}~\bibnamefont {Karthik}}, \bibinfo {author} {\bibfnamefont
  {E.}~\bibnamefont {Laermann}}, \bibinfo {author} {\bibfnamefont
  {A.}~\bibnamefont {Lahiri}}, \bibinfo {author} {\bibfnamefont
  {R.}~\bibnamefont {Larsen}}, \bibinfo {author} {\bibfnamefont {S.-T.}\
  \bibnamefont {Li}}, \bibinfo {author} {\bibfnamefont {S.}~\bibnamefont
  {Mukherjee}}, \bibinfo {author} {\bibfnamefont {H.}~\bibnamefont {Ohno}},
  \bibinfo {author} {\bibfnamefont {P.}~\bibnamefont {Petreczky}}, \bibinfo
  {author} {\bibfnamefont {H.}~\bibnamefont {Sandmeyer}}, \bibinfo {author}
  {\bibfnamefont {C.}~\bibnamefont {Schmidt}}, \bibinfo {author} {\bibfnamefont
  {S.}~\bibnamefont {Sharma}}, \ and\ \bibinfo {author} {\bibfnamefont
  {P.}~\bibnamefont {Steinbrecher}},\ }\href {\doibase
  10.1016/j.physletb.2019.05.013} {\bibfield  {journal} {\bibinfo  {journal}
  {Physics Letters B}\ }\textbf {\bibinfo {volume} {795}},\ \bibinfo {pages}
  {15} (\bibinfo {year} {2019})}\BibitemShut {NoStop}%
\bibitem [{\citenamefont {Weinberg}(2013)}]{Weinberg:1996kr}%
  \BibitemOpen
  \bibfield  {author} {\bibinfo {author} {\bibfnamefont {S.}~\bibnamefont
  {Weinberg}},\ }\href@noop {} {\emph {\bibinfo {title} {{The quantum theory of
  fields. Vol. 2: Modern applications}}}}\ (\bibinfo  {publisher} {Cambridge
  University Press},\ \bibinfo {year} {2013})\BibitemShut {NoStop}%
\bibitem [{\citenamefont {Loizides}(2020)}]{Loizides:2020tey}%
  \BibitemOpen
  \bibfield  {author} {\bibinfo {author} {\bibfnamefont {C.}~\bibnamefont
  {Loizides}},\ }in\ \href@noop {} {\emph {\bibinfo {booktitle} {{28th
  International Conference on Ultrarelativistic Nucleus-Nucleus Collisions}}}}\
  (\bibinfo {year} {2020})\ \Eprint {http://arxiv.org/abs/2007.00710}
  {arXiv:2007.00710 [nucl-ex]} \BibitemShut {NoStop}%
\bibitem [{\citenamefont {Senger}(2020)}]{Senger2020}%
  \BibitemOpen
  \bibfield  {author} {\bibinfo {author} {\bibfnamefont {P.}~\bibnamefont
  {Senger}},\ }in\ \href {\doibase 10.7566/jpscp.32.010092} {\emph {\bibinfo
  {booktitle} {Proceedings of 13th International Conference on Nucleus-Nucleus
  Collisions}}}\ (\bibinfo  {publisher} {Journal of the Physical Society of
  Japan},\ \bibinfo {year} {2020})\BibitemShut {NoStop}%
\bibitem [{\citenamefont {Harrison}\ \emph {et~al.}(2003)\citenamefont
  {Harrison}, \citenamefont {Ludlam},\ and\ \citenamefont
  {Ozaki}}]{Harrison2003}%
  \BibitemOpen
  \bibfield  {author} {\bibinfo {author} {\bibfnamefont {M.}~\bibnamefont
  {Harrison}}, \bibinfo {author} {\bibfnamefont {T.}~\bibnamefont {Ludlam}}, \
  and\ \bibinfo {author} {\bibfnamefont {S.}~\bibnamefont {Ozaki}},\ }\href
  {\doibase 10.1016/s0168-9002(02)01937-x} {\bibfield  {journal} {\bibinfo
  {journal} {Nuclear Instruments and Methods in Physics Research Section A:
  Accelerators, Spectrometers, Detectors and Associated Equipment}\ }\textbf
  {\bibinfo {volume} {499}},\ \bibinfo {pages} {235} (\bibinfo {year}
  {2003})}\BibitemShut {NoStop}%
\bibitem [{\citenamefont {Aggarwal}\ \emph {et~al.}(2010)\citenamefont
  {Aggarwal} \emph {et~al.}}]{Aggarwal:2010cw}%
  \BibitemOpen
  \bibfield  {author} {\bibinfo {author} {\bibfnamefont {M.}~\bibnamefont
  {Aggarwal}} \emph {et~al.} (\bibinfo {collaboration} {STAR}),\ }\href@noop {}
  {\  (\bibinfo {year} {2010})},\ \Eprint {http://arxiv.org/abs/1007.2613}
  {arXiv:1007.2613 [nucl-ex]} \BibitemShut {NoStop}%
\bibitem [{\citenamefont {Dainese}\ \emph {et~al.}(2017)\citenamefont {Dainese}
  \emph {et~al.}}]{Dainese:2016gch}%
  \BibitemOpen
  \bibfield  {author} {\bibinfo {author} {\bibfnamefont {A.}~\bibnamefont
  {Dainese}} \emph {et~al.},\ }\href {\doibase 10.23731/CYRM-2017-003.635}
  {\bibfield  {journal} {\bibinfo  {journal} {CERN Yellow Rep.}\ ,\ \bibinfo
  {pages} {635}} (\bibinfo {year} {2017})},\ \Eprint
  {http://arxiv.org/abs/1605.01389} {arXiv:1605.01389 [hep-ph]} \BibitemShut
  {NoStop}%
\end{thebibliography}%
\end{document}